\documentclass{eas}
\usepackage{graphicx}
\usepackage{url}
\usepackage[colorlinks, citecolor=blue]{hyperref}
\newcommand{\apj}{ApJ}

\newcommand{\aap}{A{\&}A}

\newcommand{\solphys}{Solar Phys.}

\newcommand{\ssr}{Space Sci. Rev.}

\begin{document}


\title{CME development in the corona and interplanetary medium: A multi-wavelength approach} 
\author{M. Pick} 
\address{Observatoire de Paris, LESIA, Place Jules Janssen, 92195 Meudon Cedex, France}
\author{B. Kliem} 
\address{University of Potsdam, Institute of Physics and Astronomy, 14476 Potsdam, Germany}
\secondaddress{University College London, Mullard Space Science Laboratory, Holmbury St.~Mary, UK}
\runningtitle{Pick \& Kliem: Multi-wavelength approach to CME development}
\begin{abstract}
This review focuses on the so called three-part CMEs which essentially represent the standard picture of a CME eruption. It is shown how the multi-wavelength observations obtained in the last dec\-ade, especially those with high cadence, have validated the early models and contributed to their evolution. These observations  cover a broad spectral range including the EUV, white-light, and radio domains.
\end{abstract}
\maketitle
\section{Introduction}

Since their discovery (Tousey \cite{Tousey}), coronal mass ejections (CMEs) have been extensively studied using ground-based and space-borne coronagraph observations that have enabled to analyze their basic properties. CMEs are large-scale magnetic structures which involve the expulsion of a large amount of plasma ($\sim$\,10$^{13}$--10$^{16}$~g; Vourlidas \cite{Vourlidas2010}) from the corona into the solar wind with velocities in the range $\sim$\,100--3000~km\,s$^{-1}$ (e.g., Yashiro \cite{Yashiro2004}). CMEs are often associated with eruptive prominences, or with the disappearance of filaments, and with flares. Coronagraph observations combined with data from other instruments have made possible to continuously follow the detailed progression of CMEs.
Understanding the physical processes that generate them is strongly facilitated by coordinated multi-wavelength observations. The ultimate goal is to match theory and modeling efforts with observations. This overview will mainly focus on observational aspects and will show that they strongly support  the existing models.

 \section{Kinematic evolution of CMEs}

Based on the velocity profiles, the kinematic evolution of CMEs undergoes three phases: a gradual evolution, a fast acceleration, and a propagation phase (Zhang \etal\ \cite{Zhang2001}, \cite{Zhang2004}). Most of the events reach their peak acceleration at low coronal altitudes (e.g., Gallagher \etal\
\cite{Gallagher2003}; Temmer \etal\ \cite{Temmer2008}), typically at heights below $0.5~R_\odot$ (Bein \etal\ \cite{Bein2011}). When associated with a flare, a temporal correlation both between CME velocity and soft X-ray flux and between their derivatives often exists, especially for fast CMEs (Zhang \etal\ \cite{Zhang2004}). Mari{\v c}i{\'c} \etal\ (\cite{Maricic2007}) found that 75\% of these events show this correlation while 25\% do not. The correlation indicates that the CME acceleration and the flare particle acceleration are strongly coupled. The gradual evolution
exhibits a slow rise of the structure which is about to erupt; the rise is often traced by a filament or prominence. The CME velocity changes only very gradually in the propagation phase through the outer corona and solar wind, primarily under the influence of aerodynamic drag.

\begin{figure}
\centerline{\includegraphics[width=.9\textwidth]{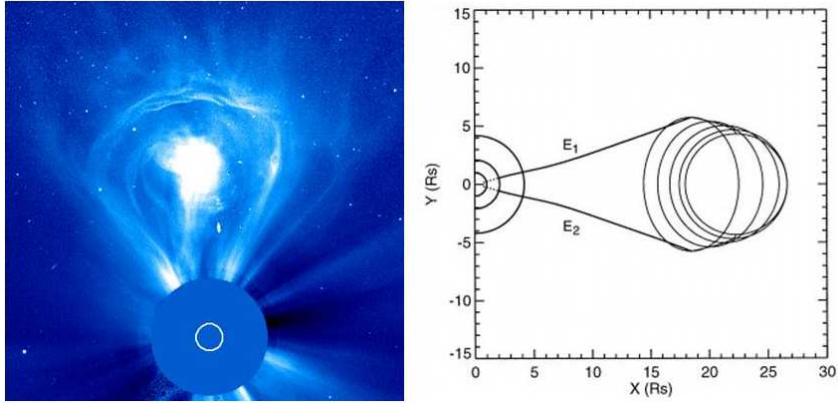}}
\caption{\textsl{Left panel:} A three-part CME showing prominence material, cavity, and an outer bright front. \textsl{Right panel:} 2D projection of the flux rope model (Chen \etal\ \cite{Chen1997}).}
\label{cme3part} 
\end{figure} 

\section{White-light coronagraphic observations: The three-part CME structure flux ropes}

	CMEs have been frequently observed in white light coronagraph images as having a so called three-part structure, consisting of a bright rim surrounding a dark void which contains a bright core (Illing \& Hundhausen \cite{Illing1985}); see Figure~\ref{cme3part}, left panel. Chen \etal\ (\cite{Chen1997}) showed that the \textsl{SOHO}/LASCO observations are consistent with a two-dimensional projection of a three-dimensional magnetic flux rope. The cavity seen in white light can be interpreted as the cross section of an expanded flux rope (Fig.~\ref{cme3part}, right panel). 

 	Beside these three-part CME structures, concave-outward V features have been frequently observed in the \textsl{SOHO}/LASCO coronagraph images (e.g., Dere \etal\ \cite{Dere1999}; St.~Cyr \etal\ \cite{StCyr2000}) and were interpreted as the sunward side of a three-dimensional helical flux rope viewed along the rope axis: see the next section and for example R{\'e}gnier \etal\ (\cite{Regnier2011}) for a detailed multi-wavelength observation of such a structure by \textsl{SDO}, which strongly supports the flux rope interpretation.

	Cremades \etal\ (\cite{Cremades2004}) showed that the projected white light structure of a three-part CME  will depend primarily on the orientation and position of the associated photospheric inversion line.

\begin{figure}
\centerline{\includegraphics[width=.9\textwidth]{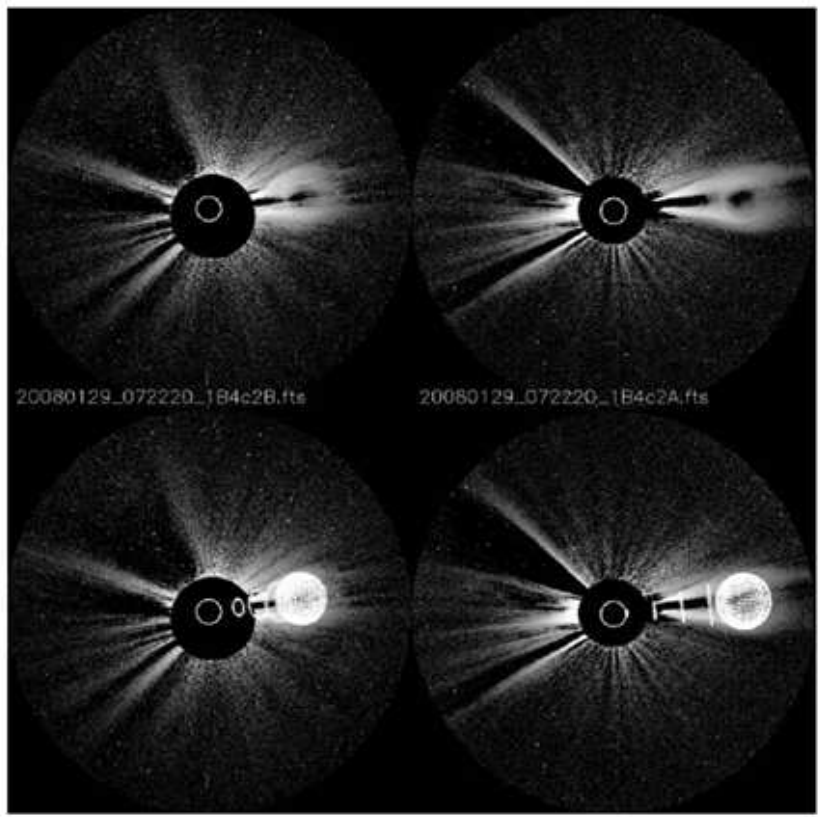}}
\caption{Example of the geometrical model fit for the event of 29 January 2008 observed at 7:20~UT in the \textsl{STEREO} coronagraphs COR2 B and A (top row). The bottom row shows the same images with the model fit overlaid (Thernisien \etal\ \cite{Thernisien2009}).}
\label{FRthernisien}
\end{figure} 
 	
	More recently, Thernisien \etal\ (\cite{Thernisien06}, \cite{Thernisien2009}) developed a 
geometric flux rope model, the gradual cylindrical shell model (GCS), and fitted it to \textsl{STEREO}/SEC\-CHI coronagraph observations of CMEs. They were able to reproduce the CME morphology for a large number of events. The flux rope orientations determined in this way revealed a deflection and/or rotation of the structure relative to the position and orientation of the source region in most cases (Fig.~\ref{FRthernisien}). 

\begin{figure}
\centerline{\includegraphics[width=.9\textwidth]{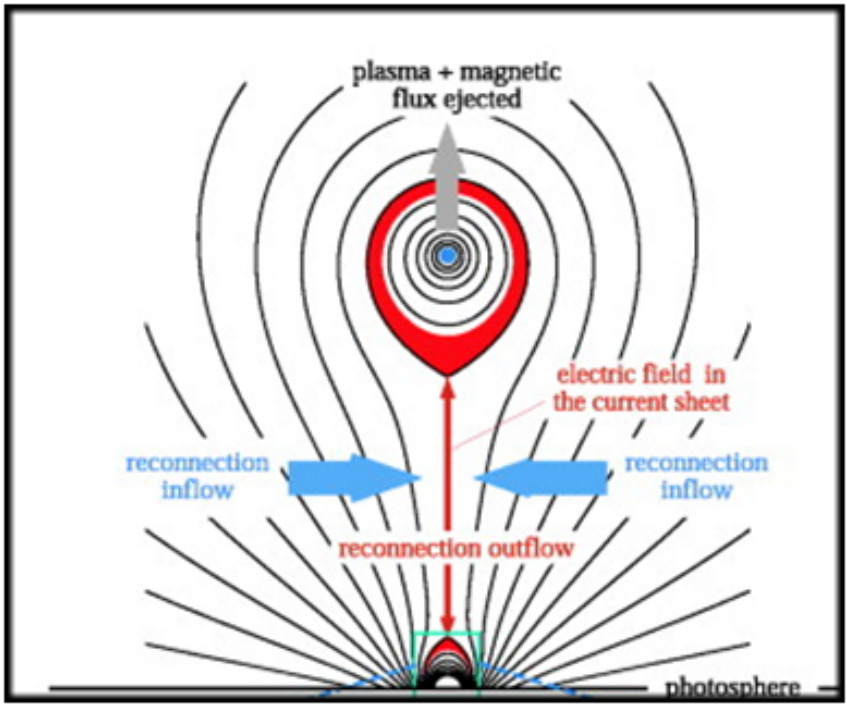}}
\caption{Sketch of the flux-rope CME model of Lin \& Forbes (\cite{Lin2000}), adapted by Lin \etal\ (\cite{Lin2004}), showing the eruption of the flux rope, the current sheet formed behind it, and the postflare/CME loops below, as well as the 
flows associated with the reconnection.}
\label{cmemodel}
\end{figure}

	The standard picture of a CME eruption emerged from the model proposed by Lin \& Forbes (\cite{Lin2000}) (Fig.~\ref{cmemodel}): An initially closed and stressed magnetic configuration overlying a photospheric polarity inversion line becomes unstable and erupts. Magnetic field lines are then stretched by the eruption and a current sheet (CS) is formed between the inversion line and the erupting flux rope. Magnetic reconnection occurs along this CS, first at low altitudes then at progressively higher ones (Forbes \& Acton \cite{Forbes1996}), producing the often associated flare and also explaining the formation of post-eruption loops behind the CS.

This model is the synthesis of the loss-of-equilibrium model for the upward acceleration of the filament and CME (van Tend \& Kuperus \cite{vanTend1978}; Forbes \& Isenberg \cite{Forbes1991}) and the standard (\ie, reconnection) model of eruptive flares, also known as the CSHKP model (Carmichael \cite{Carmichael1964}; Sturrock \cite{Sturrock1966}; Hirayama \cite{Hirayama1974}; Kopp \& Pneuman \cite{Kopp1976}). It includes two stages. The source region, which is supposed to already contain a flux rope, first stores free magnetic energy in a quasi-static evolution, driven by slow changes of the photospheric field. The resulting inflation of the coronal field is observed as the slow rise. When a critical point is reached, the flux rope loses equilibrium and is rapidly accelerated upwards by the Lorentz force of the current flowing in the rope (onset of CME or filament upward ejection). This is coupled with the flare reconnection in the vertical CS, which reduces the tension of the overlying field. The model is now strongly supported by numerous observations made in different wavelength domains (see Benz \cite{Benz2008} and Fletcher \etal\ \cite{Fletcher2011} for the flare observations).
One of the major open questions of CME research is whether the MHD instability of the current-carrying flux rope or the reconnection in the CS underneath is the main driver of the eruption as a whole.

\section{Multi-wavelength CME observations and comparison with theoretical predictions}

	Because magnetic reconnection may be occurring in the CS 
 and enabling energy release, the presence of a thin spike of high-temperature material behind a CME would match the expectation of the standard model. In recent years, many observational evidences of such CS have been
found in white light coronagraph observations, X-ray images, coronal
UV spectra, EUV images, radio spectra, and radio images.

\begin{figure}
\centerline{\includegraphics[width=.9\textwidth]{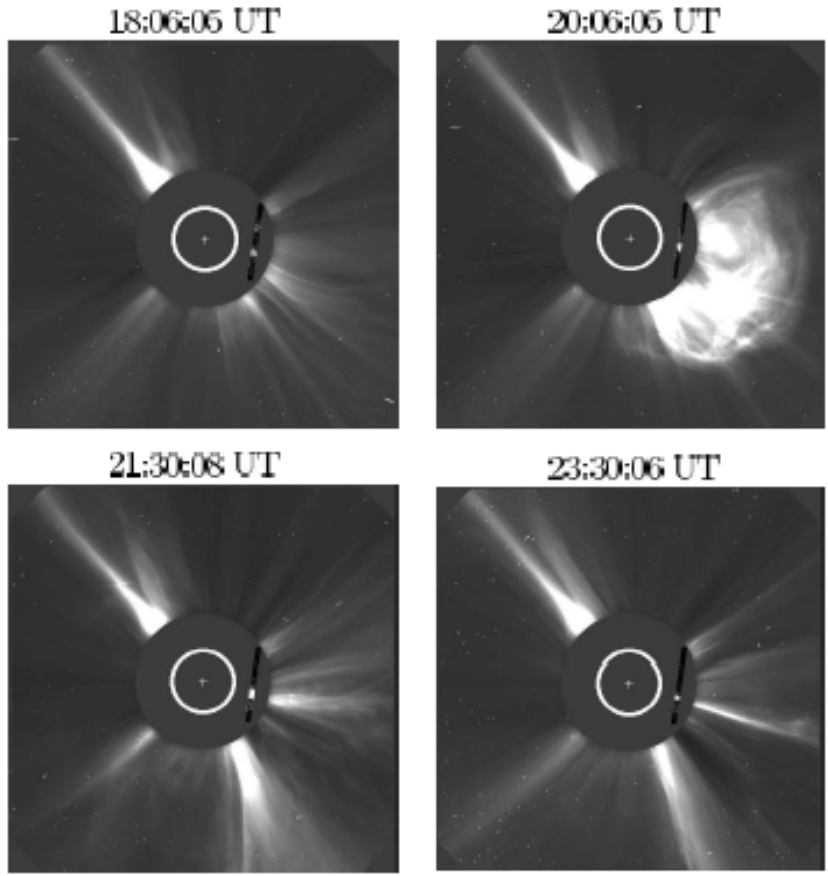}}
\caption{\textsl{SOHO}/LASCO C2 images 
showing the evolution of the events from the pre-CME corona to the narrow long CS feature. The images shown inside the occulter disk are the intensity distribution of the Fe~XVIII line along the UVCS slit. The brightest narrow spot is very well aligned with the CS seen by LASCO (Ciaravella \& Raymond \cite{Ciaravella2008}).}
\label{UVCS}
\end{figure} 

In the UV domain, the \textsl{SOHO}/UVCS coronagraph has observed several CME events exhibiting high-temperature emissions from the Fe~XVIII 974~\AA\ line at heliocentric heights of 1.5--1.7~$R_\odot$ which lie along the line connecting the eruptive CME and the associated post-CME loops (Fig.~\ref{UVCS}). The location and the timing of these emissions strongly support the interpretation 
that a post-CME CS formed (Ciaravella \& Raymond \cite{Ciaravella2008}; Ko \etal\ \cite{Ko2010}). 

In white-light, Vr{\v s}nak \etal\ (\cite{Vrsnak09}) analyzed the morphology and density structure of rays observed by the \textsl{SOHO}/LASCO C2 coronagraph in the aftermath of CMEs. The most common form of activity is characterized by outflows along the rays, and sometimes also by inflows. The authors concluded that the main cause of density excess in these rays is the upward transport of the dense plasma by the reconnection outflow
in the CS formed in the wake of CMEs. 

\subsection{EUV Observations from the Solar Dynamical Observatory}

\begin{figure}
\centerline{\includegraphics[width=.9\textwidth]{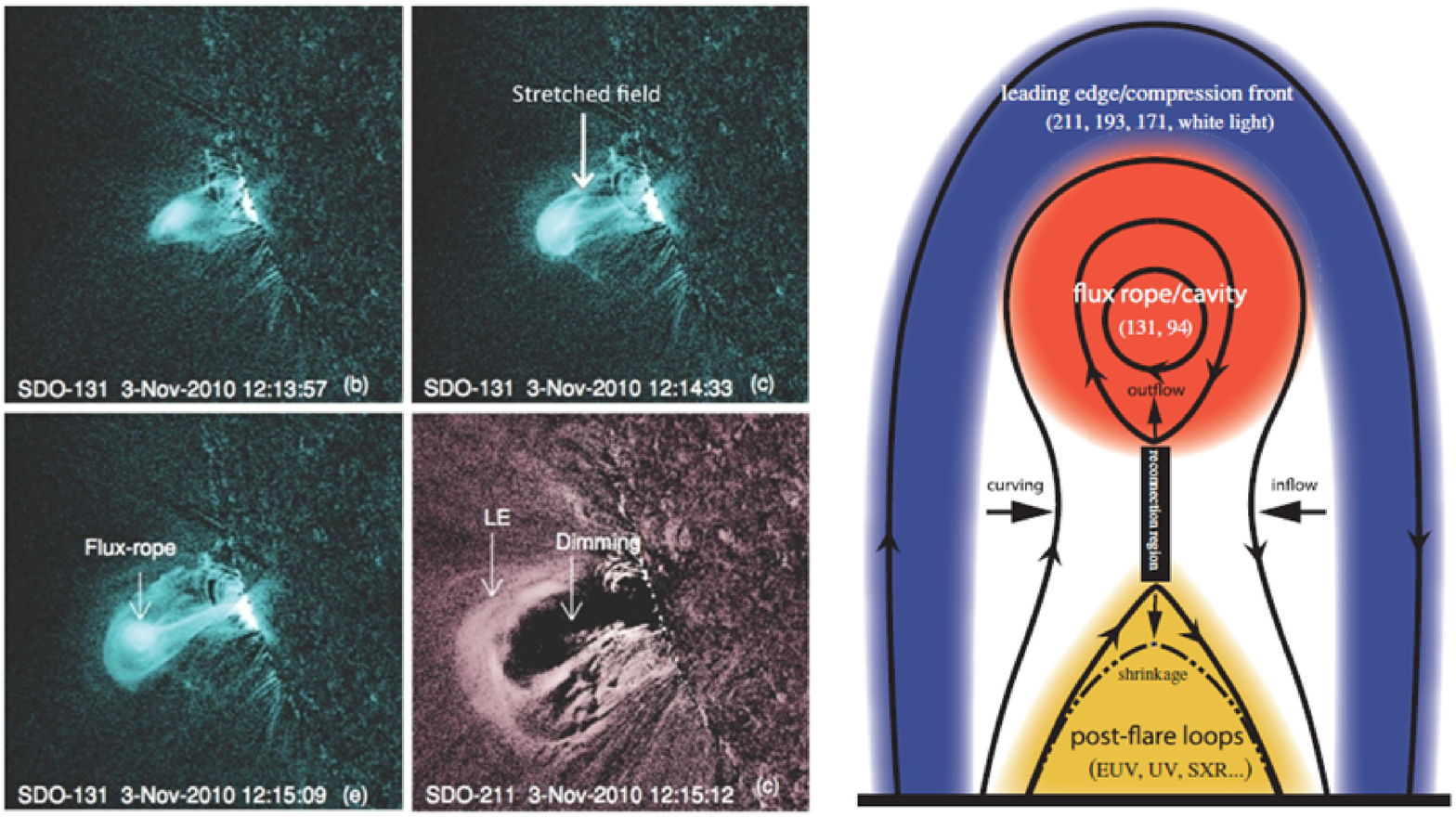}}
\caption{LEFT: \textsl{SDO}/AIA base-difference images of the solar eruption on 2010 November 3 at 131~\AA\ ($\sim$~11~MK) (\textsl{upper panel and lower-left panel}),  and at 211~\AA\ ($\sim$~2~MK) (\textsl{lower-right panel}).  Leading edge and dimming features are indicated by 
arrows. RIGHT: CME structure as seen in AIA multiple temperature bands 
(from Cheng \etal\ \cite{Cheng2011}).}
\label{SDO}
\end{figure} 
The most recent and fascinating results arise from the Atmospheric Imaging Assembly (AIA) on board of the \textsl{Solar Dynamic Observatory (SDO)}. This experiment has the capability of high cadence and multi-temperature observations. Figure~\ref{SDO}, left, adapted from Cheng \etal\ (\cite{Cheng2011}), shows a few base-difference images in the 131~\AA\ bandpass (dominated by 11~MK plasma) of the solar eruption on 2010
Nov~3 which was associated with a limb CME detected by \textsl{SOHO}/LASCO. A blob of hot plasma appeared first and started to push its overlying magnetic field upward. The overlying field lines seem to be stretched up continuously. Below the blob, there appeared a Y-type magnetic configuration with a bright thin line extending downward, which is consistent with the development of a CS. In addition, the shrinkage of magnetic field lines underneath the CS indicates the ongoing process of magnetic reconnection. The plasma blob likely corresponds to a growing flux rope. Simultaneously, a cavity with diffuse density enhancement at the edge is seen at typical coronal temperatures ($T\sim0.6\mbox{--}2$~MK). For the first time, the multi-temperature structure of the CME has been analyzed 
(see Fig.~\ref{SDO} right).

\begin{figure}
\centerline{\includegraphics[width=\textwidth]{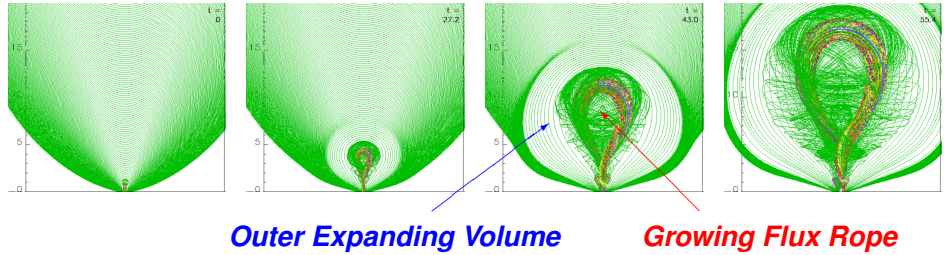}}
\caption{MHD simulation of cavity formation around an erupting flux rope. Rainbow-colored field lines show the core of the flux rope. Green field lines show the ambient field, a progressively larger fraction of which becomes part of the rope, due to reconnection in the CS under the rope (from Kliem \etal, in preparation).}
\label{simulation}
\end{figure}

\begin{figure}
\centerline{\includegraphics[width=.9\textwidth]{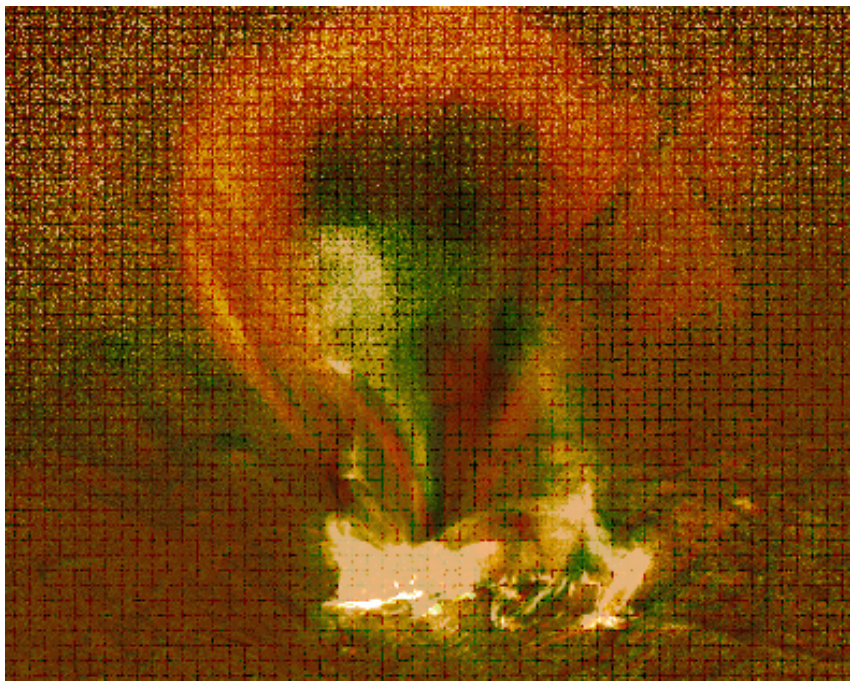}}
\caption{Rapidly forming cavity around a CME flux rope. Overlay of \textsl{SDO}/AIA images in the 131~{\AA} ($\sim11$~MK, green) and 171~{\AA} ($\sim0.6$~MK, red) channels (Kliem \etal, in~prep.).}
\label{FR-cav}
\end{figure}

The high-cadence EUV observations by \textsl{STEREO} and \textsl{SDO} also yield insight how the forming CME expands the ambient field. This process rapidly forms a cavity 
which surrounds the plasma in the hot CS and flux rope 
(Figs.~\ref{SDO} and \ref{FR-cav}). A cavity exists around some prominences prior to their eruption, especially in quiescent prominences (Gibson \etal\ \cite{Gibson2006}), but many events lack such signatures, especially the cavities forming rapidly around erupting active region filaments/prominences (e.g., Patsourakos \etal\ \cite{Patsourakos2010b}; Cheng \etal\ \cite{Cheng2011}). The growth of the cavity certainly reflects the growth of the flux rope in the CME core, due to the addition of flux by reconnection; however, initially the cavity, or ``bubble'', grows even faster than the flux rope in some events. This has led Patsourakos \etal\ (\cite{Patsourakos2010a}) to suggest an additional expansion mechanism based on ideal MHD effects (\ie, independent of reconnection). The mechanism assumes that the free energy released in the eruption is contained in the current of a flux rope that exists already prior to the onset of the CME. When the flux rope rises, the current through the rope decreases, powering the eruption; the decrease is approximately inversely proportional to the length of the rope. As a consequence, the azimuthal (poloidal) field component in and around the flux rope must also decrease. Since the total poloidal flux in the system is not changed by the rise of the rope, the flux surfaces must move away from the center of the flux rope to reduce the strength of the poloidal field component. In other words, the poloidal flux in and around the flux rope must expand, forming a cavity (or deepening the cavity if it existed already before the eruption). The numerical simulation of an erupting flux rope displayed in Figure~\ref{simulation} clearly exhibits both mechanisms of cavity formation and expansion.
The relative contributions of the two mechanisms of cavity growth vary from event to event. The flux expansion of the ideal MHD mechanism is driven by the poloidal field component, it has to work against the toroidal (shear field) component. If the latter is strong, this part of the expansion will be slowed and weakened. The pressure has a similar influence if the plasma beta is not very small (about 0.1 or larger). Consequently, some events show the cavity edge quite close to the edge of the growing flux rope (Fig.~\ref{SDO}), while others show a cavity much larger than the flux rope (Fig.~\ref{FR-cav}).

The cases of very rapid initial cavity expansion are of particular interest as potential sources of large-scale coronal EUV waves (also known as ``EIT Waves'') and shocks. 
It has 
been recognized that the initial expansion of the CME is the prime candidate for the formation of these phenomena. This replaces the conjecture of a flare blast wave. 
The rapid cavity expansion may eventually also help solving the puzzle why many coronal shocks, seen as Type II radio bursts, appear to be launched at the side of the expanding CME, not at its apex. The triggering of an EUV wave by a rapidly expanding CME cavity, including the formation of a shock, has very recently been demonstrated, again using multi-wavelength data from 
\textsl{SDO} combined with radio data (Cheng \etal\ \cite{Cheng2012}).

\subsection{Radio observations}

	Radio spectral and imaging observations are obtained with extremely high time resolution and sample different heights in the solar atmosphere. So they contribute significantly to our understanding of CME initiation and development as briefly summarized below. The first indications of CSs in the solar corona were provided by radio spectral observations. Kliem \etal\ (\cite{Kliem2000}) observed long series of quasi-periodic pulsations deeply modulating a continuum in the {$\sim(1\mbox{--}2)$~GHz range that was slowly drifting toward lower frequencies. They proposed a model in which the pulsations of the radio flux are caused by quasi-periodic particle acceleration episodes that result from a dynamic phase of magnetic reconnection in a large-scale CS (see also Karlick{\'y} \etal\ \cite{Karlicky2002}; Karlick{\'y} \cite{Karlicky2004}; Karlick{\'y} \& B{\'a}rta \cite{Karlicky2011}}). Such breakup of the CS into filamentary structures can cascade to the smallest scales (B{\'a}rta \etal\ \cite{Barta2011}). The possible transition to a turbulent regime of reconnection is currently of high interest even beyond the solar context (e.g., Lazarian \& Opher \cite{Lazarian2009}; Daughton \etal\ \cite{Daughton2011}).

\subsubsection{Radio-imaging, X-ray and EUV observations}

\begin{figure}
\centerline{\includegraphics[width=\textwidth]{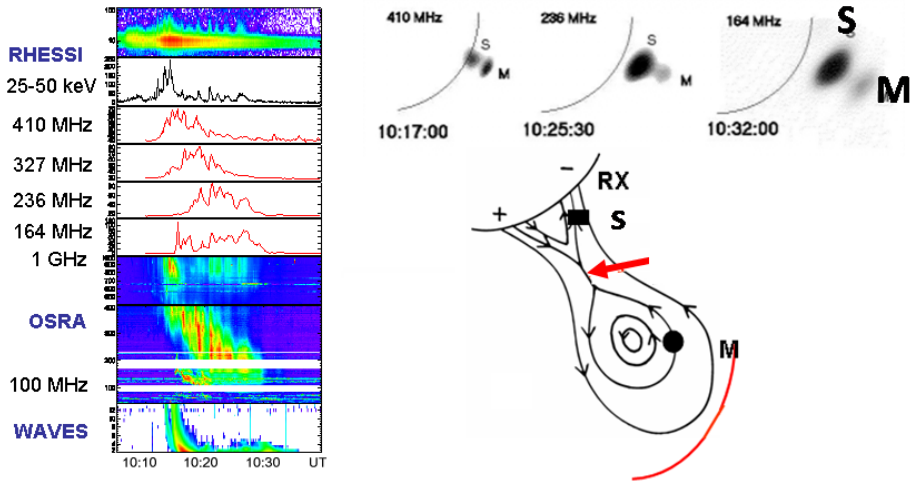}}
\caption{Radio and X-ray signatures of magnetic reconnection behind an ejected flux rope on 02 June 2002. \textsl{Left panel}: Comparison between the photon histories measured by \textsl{RHESSI}, the flux evolution measured at four frequencies by the NRH and the spectral evolution measured by OSRA and by \textsl{STEREO}/WAVES. \textsl{Right panel, top}: Images of the Nan\c cay Radioheliograph (NRH) at 410, 236, and 164 MHz, showing the quasi-stationary sources (S) and the moving sources (M). The event is close to the solar limb (curved line). \textsl{Right panel, bottom}: Two-dimensional sketch of the magnetic configuration involved in the  eruption. A twisted flux rope erupts, driving magnetic reconnection behind it  (red arrow). The particles accelerated in the reconnection region propagate along the reconnected field lines, giving rise to the observed hard X-rays (RX) and the main radio sources (S and M). A shock is propagating at the front edge of the flux rope (red curve) (from Pick \etal\ \cite{Pick2005}).}
\label{cme-radio}
\end{figure}

Pick \etal\ (\cite{Pick2005}) traced the dynamical evolution of the reconnecting CS behind an ejected flux rope and provided an upper estimate of the CS length from the position of the observed pair of radio sources, consisting of an almost stationary and a rapidly moving source (see Fig.~\ref{cme-radio}). Later, Aurass \etal\ (\cite{Aurass2009}) provided diagnostics of the presence of a CS in the aftermath of a CME both with X-ray and radio spectral observations, and Benz \etal\ (\cite{Benz2011}) imaged the CS in radio, showing that it extended above the temporally correlated, largely thermal coronal X-ray source. Finally, Huang \etal\ (\cite{Huang2011}) demonstrated that joint imaging radio and EUV observations can trace the extent and orientation of the flux rope and its interaction with the surrounding magnetic field. This allows to characterize in space and time the processes involved in the CME launch.

Bastian \etal\ (\cite{Bastian2001}) first reported the existence of an ensemble of expanding loops that were imaged in radio by the NRH and were located behind the front of the white light CME on 1998 April 20. The faint emission of these loops, named \textit{radio CME},  was attributed to incoherent synchrotron radiation from 0.5~MeV electrons spiraling within a magnetic field ranging from 0.1 to a few Gauss. Maia \etal\  (\cite{Maia2007}) identified another radio CME on 2001 April 15 which was one of the largest one of this solar cycle. A recent study established that the radio CME corresponds to the flux rope seen in white light and that its extrapolated center coincides with the center of the flux rope cavity (D{\'e}moulin \etal\ \cite{Demoulin2012}). The CS behind the flux rope was also imaged in radio.

\begin{figure}
\centerline{\includegraphics[width=\textwidth]{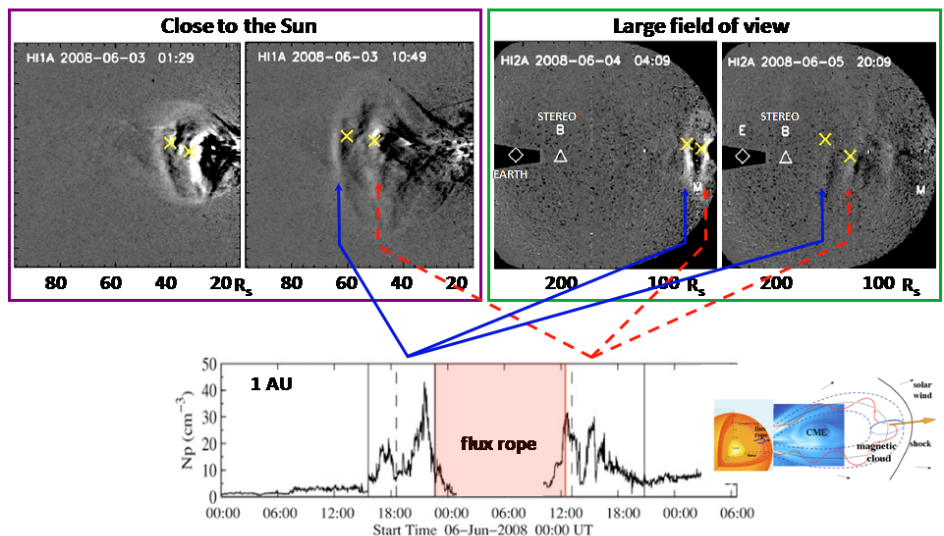}}
\caption{Linking remote imagery to in-situ signatures. TOP: Evolution of the CME in \textsl{STEREO-A}/HI1 and \textsl{STEREO-B}/HI1. The CME leading edge and core are given by yellow cosses. BOTTOM: \textsl{Left}: Proton density measured at 1 AU by \textsl{STEREO-B}; the dashed lines are the arrival times from the elongation fitting method for the CME leading edge (see blue lines) and CME core (see red lines) (adapted from M{\"o}stl \etal\ (\cite{Mostl2009}), courtesy of P. D{\'e}moulin). \textsl{Right}: Schematic showing three evolutionary steps. First, the crossing of the convective zone by a flux rope. Second, the launch of a CME. Finally, depending on the speed and launch direction, the CME can be detected a few days later in the IP medium as as a magnetic cloud or more generally as an ICME. The CME image is from \textsl{SOHO}/LASCO (from D{\'e}moulin \cite{Demoulin2010aipc}).}
\label{cme-interplanetary}
\end{figure}   

\section{Relationship between CMEs and interplanetary coronal mass ejections (ICMEs)}

The \textsl{STEREO} HI1 and H2 imagers have provided the opportunity to trace the evolution of CMEs to 1~AU and beyond and thus to investigate for the first time the relationship with their heliospheric counterpart in the whole inner heliosphere. The observations suggest that many CMEs are still connected to the Sun at 60-80~$R_\odot$ and that the same basic structure is often preserved as the CME propagates from the corona into the heliosphere (e.g., Harrison \etal\ \cite{Harrison2010}). Figure~\ref{cme-interplanetary}, upper panel shows four \textsl{STEREO}-A images of a CME which was associated with a magnetic cloud (MC). In the HI1A and HI2A images, the CME leading edge (LE) and core, indicated in the figure by yellow crosses, show an arc-like shape, typical of a CME viewed orthogonal to its axis of symmetry. M{\"o}stl \etal\ (\cite{Mostl2009}) linked the remote observations of this CME to the MC plasma and magnetic field data measured by \textsl{STEREO}-B at 1~AU. Figure~\ref{cme-interplanetary} shows that the three-part structure of the CME may be plausibly related to the in-situ data and that the CME white-light flux rope corresponds to the magnetic flux rope (MFR) measured in situ.


\section{Conclusion}

In this brief review we  have focused almost exclusively on the \textit{three-part CMEs} and how the multi-wavelength observations, in particular when obtained with high cadence, have validated the early models and contributed to their evolution. It  must be recalled, however, that fast flare/CME events often have a much more complex development. A detailed presentation of these CME events is beyond the scope of this review. We shall only mention that they  are observed to start with a relatively small dimension and then reach their full extension by the rapid expansion of the cavity and by successive interactions with the surrounding magnetic structures. NRH observations show that they  can cover a large portion of the Sun within typically 10~min or even less (for a review see Pick \etal\ \cite{Pick2008}). Other important topics that could not or only briefly be covered here are i) the initiation phase preceeding the onset of a CME; ii) the association between CMEs, flares, EIT/EUV waves, and coronal and interplanetary shocks (Type II radio bursts), and iii) the role of dimmings (transient coronal holes) in the dynamics of CMEs.

The unprecedented observational capabilities now available (imaging and spectroscopy
from radio to hard X-rays; stereoscopy; high spatial resolution and cadence; imaging from the Sun to 1~AU and beyond), combined with similar progress in the numerical modeling, will undoubtedly stimulate further discoveries and deeper understanding of CMEs, flares and their associated phenomena. Some of the most challenging directions in the future research will be : i) the mutual feedback between CMEs and flares;
 ii) the coupling of the smallest scales (reconnection, particle acceleration) with the largest ones (CME, large-scale CS, sympathetic eruptions); and iii) the connection with photospheric and subphotospheric phenomena (triggering by flux emergence, back-reaction forming sun quakes).


\end{document}